\documentclass[10pt,notitlepage,a4paper,aps,prd,onecolumn,superscriptaddress,nofootinbib,groupedaddress]{revtex4-1}

\usepackage{amsmath}         
\usepackage{amsfonts}        
\usepackage{amssymb}         
\usepackage{amsthm}          
\usepackage{empheq}          

\usepackage[usenames,dvipsnames]{color}           
\usepackage{ragged2e}        
\usepackage{titlesec}        

\usepackage{tipa}            
\newcommand{\thorn}{\mbox{\textthorn}}  

\usepackage[utf8]{inputenc} 
\allowdisplaybreaks          

\usepackage{enumitem}        

\usepackage{tikz}            
\usetikzlibrary{shapes.geometric,arrows,positioning}  

\newcommand\blfootnote[1]{%
  \begingroup
  \renewcommand\thefootnote{}\footnote{#1}%
  \addtocounter{footnote}{-1}%
  \endgroup
}

\titleformat{\paragraph}
  {\normalfont\normalsize\bfseries}{\theparagraph}{1em}{} 
\titlespacing*{\paragraph}
  {0pt}{3.25ex plus 1ex minus .2ex}{1.5ex plus .2ex} 

\usepackage[colorlinks=true, linkcolor=blue, citecolor=blue, filecolor=blue, urlcolor=blue,hyperfootnotes=true]{hyperref}        

\begin{document}
\title{The Smarr formula within the Geroch-Held-Penrose formalism}

\author{Alberto Guilabert}
\email{Corresponding author: alberto.guilabert@ua.es}
\affiliation{Fundacion Humanismo y Ciencia, Guzmán el Bueno, 66, 28015 Madrid, Spain.}
\affiliation{Departamento de F\'{\i}sica Aplicada, Universidad de Alicante, Campus de San Vicente del Raspeig, E-03690 Alicante, Spain.}

\author{Pedro Bargueño}
\email{pedro.bargueno@ua.es}
\affiliation{Departamento de F\'{\i}sica Aplicada, Universidad de Alicante, Campus de San Vicente del Raspeig, E-03690 Alicante, Spain.}

\begin{abstract}
The connection between classical thermodynamics and black hole horizons is a fundamental topic in gravitational physics, offering a potential pathway to understanding quantum aspects of gravity. However, while black hole mechanics exhibits well-known thermodynamic parallels, a rigorous geometric interpretation of thermodynamic variables directly from the field equations warrants further research. In this manuscript, we present a thermodynamic formulation of the field equations through the decomposition of the Riemann tensor, employing the Geroch-Held-Penrose (GHP) formalism, to clarify a strong correspondence between black hole thermodynamic variables and geometrical quantities derived from horizon geometry. Our analysis reveals an intrinsic connection between the Penrose and Rindler $K$-curvature and the Smarr relation, motivating a revised definition of both trapping gravity and black hole internal energy. Additionally, we derive through this GHP formalism the Smarr formula for the Reissner-Nördstrom black hole cointained in an AdS spacetime and we explore the implications of this relationship for black holes with exotic topologies and in the context of extended theories, exemplified by $f(R)$ gravity. We also briefly analyze this relation in more general configurations, such as Taub-NUT spacetimes. These findings suggest a deeper geometrical basis for black hole thermodynamics, potentially advancing our understanding of gravitational energy, horizon entropy, and their significance within quantum gravity frameworks.
\end{abstract}

\blfootnote{}

\maketitle

\tableofcontents


\section{Introduction}\label{sec:intro}

After Hawking's area theorem for black holes \cite{Hawking1972}, which states that black hole surface cannot decrease\footnote{This initial proof assumed asymptotic flatness, the weak energy condition and cosmic censorship. These hypothesis can be restrictive when considering aspects of quantum gravity. Some generalizations of this law have been developed \cite{Woolgar1999,Kontou2024}.} as a result of any classical process, a new topic in gravitation was developed which tries to study the connections between thermodynamics (possibly from a statistical mechanics perspective) and black hole horizons.

At the same time, Christodoulou showed that black holes possess an irreducible mass which is strongly related to the black hole surface area \cite{Christodoulou1970,Christodoulou1971}. This includes even charged Kerr black holes in the context of General Relativity (GR). Building on this, the Penrose inequality conjectures a more general geometric bound: the total ADM mass of a spacetime containing a black hole cannot be less than the irreducible mass associated with the event horizon's area. Actually, generalized versions of this inequality are still being developed \cite{Harvie2024}.

The black hole area theorem, the existence of an irreducible mass and their similarity with the fundamental notion of classical entropy motivated Bekenstein to obtain an expression for black hole entropy proportional to the area of the horizon \cite{Bekenstein1973}. More generally, the black hole entropy was found later to be a Noether charge of the Hamiltonian due to the diffeomorphism invariance property of any\footnote{We are assuming that the general covariance principle is necessary to formulate any suitable theory of gravity.} theory of gravity \cite{Wald1993}. This result extended the study of black hole thermodynamics to more general theories of gravity \cite{Faraoni2010,Padmanabhan2010}. In particular, the thermodynamics of $f(R)$ gravity has been extensively studied, including various formulations of this theory \cite{Bamba2010, Peralta2020}. This conserved charge coincides with the Bekenstein-Hawking entropy in GR. Furthermore, quantum corrections to entropy have been studied in different contexts to give a physical interpretation to the entropy of black holes \cite{Susskind1993}.

At this point, these results showed the existence of a direct analogy with the second law of thermodynamics, but replacing the entropy by the black hole area. Then, we can see the spacetime as a closed\footnote{As we are not aware of any energy or matter exchange with a \emph{surrounding} environment.} thermodynamic system where time moves in a direction such that the area of a black hole increases.

This rise in entropy corresponds to an increase in energy that cannot be converted into work. Similarly, the irreducible mass of a black hole represents the portion of energy that cannot be extracted. Hence, this irreducible mass can be interpreted as the energy that cannot be transformed into usable work.

The entropy associated with an event horizon strongly suggests the existence of its thermodynamic conjugated variable analog to the temperature. Based on this and considering quantum mechanical effects occurring at the horizon, Hawking discovered that black holes emit thermal radiation \cite{Hawking1975}. More generally, a temperature can be associated to any bifurcation horizon \cite{Lee1986} or even to a generic null surface \cite{Dalui2021}.

Black hole thermodynamics, which emerged from these initial results, establishes formal analogies between the behavior of black holes and the fundamental laws of thermodynamics. The pioneering work of Bekenstein and Hawking led to the reformulation of the four laws of black hole mechanics \cite{Bardeen1973} in terms of the classical laws of thermodynamics.

The zeroth law establishes that the surface gravity of a stationary black hole is constant across its horizon, resembling the constancy of temperature in thermal equilibrium. The first law links changes in a black hole mass, angular momentum, and charge to changes in area, thus establishing an analogy with the thermodynamic relation between energy, temperature, and entropy.

The second law, as derived from Hawking’s area theorem, suggests that the area of a black hole horizon never decreases, much like the entropy in a closed thermodynamic system, leading to the concept of black hole entropy being proportional to the area of the horizon. Finally, a third law was proposed, which posits that it is impossible to reduce a black hole’s surface gravity to zero through any physical process, analogous to the thermodynamic third law.

The study of black hole thermodynamics has also motivated a broader look at the thermodynamic properties of horizons in non-black hole contexts, such as in accelerating or cosmological horizons \cite{Appels2016,Gibbons1977b}, further suggesting that thermodynamic-like descriptions may apply to a wide range of horizon-like structures in gravitational settings.

A finding problem in the study of thermodynamics is that event horizons are global objects defined on asymptotically flat spacetimes. In contrast, our universe might not even be asymptotically flat considering a non-zero cosmological constant, meaning that event horizons might not exist in practical terms. Even if the universe were asymptotically flat, finite observers would be unable to detect or confirm such a global property.

Within this context, Hayward gave a general quasi-local definition of a black hole in terms of future outer trapping horizons \cite{Hayward1994d}. These trapping horizons are hypersurfaces constructed by a specific foliation of future outer marginal surfaces, which are Riemannian 2-surfaces where a null expansion vanishes (these marginal surfaces are the 2-dimensional \textit{horizons} usually seen in pictures or movies). Moreover, a generalization of the surface gravity, called trapping gravity, and the related black hole dynamic laws are given, which were also translated into GHP formalism, allowing a more geometric interpretation \cite{Hayward1994}.

A key limitation is that Hayward's definition for trapping gravity (see Eq. (7) in \cite{Hayward1994}), does not coincide with the surface gravity even in the context of static and spherically symmetric spacetimes, e.g. the Reissner-Nördstrom black hole \cite{Fodor1996,Nielsen2008}. A revised definition is proposed in this manuscript based in the first law of black hole thermodynamics.

Finally, let us comment that connections between thermodynamics and gravity have been explored at a fundamental level. Jacobson derived Einstein equations as a consequence of the first law of thermodynamics \cite{Jacobson1995}, highlighting their intrinsic relationship. In addition, the field equations of $f(R)$ gravity were derived in \cite{Eling2006}, where the Clausius relation was extended to non-equilibrium thermodynamics within this framework. 

Conversely, Padmanabhan demonstrated that the first law of thermodynamics can also be derived from Einstein’s equations in certain simplified scenarios \cite{Padmanabhan2002}, underlying the deep connection between gravity and thermodynamics.

The study of black hole thermodynamics has revealed deep connections between gravity, thermodynamics, and quantum theory. However, a key challenge arises from the global nature of event horizons, which limits their physical applicability in realistic cosmological settings. Quasi-local approaches, such as Hayward’s trapping horizons, offer a more general framework, but existing definitions of trapping gravity fail to recover the expected thermodynamic properties in well-known cases. Based on Padmanabhan's approach, we propose a revised definition of trapping gravity inspired by the first law of black hole thermodynamics, aiming to provide a more consistent description within the Geroch-Held-Penrose (GHP) formalism. By formulating black hole thermodynamics in a purely geometric language, we hope to deepen our understanding of the fundamental relationship between gravity and thermodynamics and explore its broader implications.

This manuscript is organized as follows: Sec. \ref{sec:rep_padma} provides a brief overview of Padmanabhan's results connecting the Einstein field equations to the first law of black hole thermodynamics in static and spherically symmetric spacetimes. In Sec. \ref{sec:general_case} we present the GHP formulation of the field equations, analyzing the nontrivial equations depending on the underlying geometric structure of the spacetime and their relation with geometrical quantities. In Sec. \ref{sec:main_findings}, we present the key findings of this work, exploring thermodynamics within the GHP formalism for static spacetimes foliated by compact orientable 2-surfaces of constant Gaussian curvature. Eventually, in Sec. \ref{sec:final_remarks} we present some final remarks and interesting directions for future research.

\section{Padmanabhan's view of Einstein field equations as a thermodynamic relation}\label{sec:rep_padma}

Initially, the laws of black hole thermodynamics were constructed without mentioning the Einstein field equations. Surprisingly, Padmanabhan showed that the first law of black hole thermodynamics is also encoded in the field equations \cite{Padmanabhan2002}. This was done first in the context of GR for static and spherically symmetric spacetimes with line element
\begin{equation}\label{metric:static_spherical}
ds^2 = f(r) dt^2 - f(r)^{-1} dr^2 - r^2 d\Omega^2.
\end{equation}

In this case, a sufficient condition to have a non-degenerate outer trapping horizon at $r=a$ is given by $f(a)=0$ and $f'(a)\neq 0$. Then, the trapping horizon is a timelike hypersurface foliated by constant $t$ slices, which are the future outer marginal surfaces. Additionally, condition $f'(a) > 0$ is needed to ensure that the horizon is of future type, which is related to the presence of a black hole.

Metrics of the form of Eq. \eqref{metric:static_spherical} can be reduced to the Rindler metric near the horizon, where the acceleration of the Rindler observer is determined by its surface gravity. Thus, it is possible to identify the horizon temperature\footnote{We are working with geometrized units, $G=c=k_B=1$.} as
\begin{equation}\label{ec:Temperature_Padmanabhan}
T = \dfrac{\hbar f'(a)}{4\pi}.
\end{equation}

In addition, the only nonzero components of the Einstein tensor satisfy the relations $G_t^t = G_r^r$ and $G_{\theta}^{\theta} = G_{\phi}^{\phi}$. Padmanabhan focused on the thermodynamic interpretation of the radial equation, $G_r^r = T_r^r$, which evaluated at the horizon, $r = a$, can be written as
\begin{equation}\label{eq:G_r^r Padmanabhan}
\dfrac{f'(a)}{2}a-\dfrac{1}{2} = 4\pi T_r^r a^2.
\end{equation}

After multiplying by $\mathrm{d}a$ and rearranging terms, this expression leads to
\begin{equation}\label{ec:First_Law_Padmanabhan}
\dfrac{\mathrm{d}a}{2} = \dfrac{\hbar f'(a)}{4\pi} \mathrm{d}\left(\dfrac{1}{4\hbar}4\pi a^2\right) + T_r^r \mathrm{d}\left(\dfrac{4\pi}{3}a^3\right),
\end{equation}
which is identified with the first law of black hole thermodynamics after recognizing the internal energy of the black hole as the term on the left hand side, the temperature of the horizon from Eq. \eqref{ec:Temperature_Padmanabhan} and the Bekenstein-Hawking entropy $S = A/4\hbar$. That is,
\begin{equation}\label{ec:differential_first_law}
\mathrm{d}E = T\mathrm{d}S - P\mathrm{d}V,
\end{equation}
where the pressure is defined as $P=-T_r^r$. Moreover, this identification is uniquely fixed \cite{Padmanabhan2010}.

The corresponding Smarr relation can be obtained by using a scaling argument and Euler's theorem \cite{Kastor2009} to integrate Eq. \eqref{ec:differential_first_law}. Note that $E\sim r$, $S\sim r^2$ and $V\sim r^3$. Thus, the Smarr formula is given by 
\begin{eqnarray}\label{ec:Smarr_Padmanabhan}
E = 2TS - 3PV.
\end{eqnarray}

Note that this relation is not exactly the same as the Smarr mass formula for charged Kerr black holes \cite{Smarr1973}. In fact, Eq. \eqref{ec:differential_first_law} and its corresponding integrated form shown in Eq. \eqref{ec:Smarr_Padmanabhan} generalize the Smarr formula for any static and spherically symmetric black hole in GR.

For the vacuum case and without cosmological constant, $E$ coincides with the mass of the black hole and Eq. \eqref{ec:Smarr_Padmanabhan} reduces to $M = 2TS$. Alternatively, $E$ does not coincide with the mass of the black hole when considering a cosmological constant \cite{Kubizk2017}. Actually, when the pressure term is due to the presence of a negative cosmological constant, $P = -\lambda/8\pi$, the variable $E$ denotes the total internal energy of the black hole and its mass may be identified with the black hole chemical enthalpy $M = E + PV$ \cite{Cveti2011}. In such a case, the Smarr relation in Eq. \eqref{ec:Smarr_Padmanabhan} takes the form
\begin{equation}
M = 2TS-2PV.
\end{equation}

Finally, let us point out that this derivation was extended to any static spacetime for Lanczos-Lovelock gravity \cite{Kothawala2009} and for stationary axisymmetric horizons \cite{Kothawala2007}.


\section{\label{sec:general_case}GHP field equations}

As we commented before, here we will extend the thermodynamic analysis done by Padmanabhan to more general scenarios. Specifically, we will study static spacetimes foliated by compact orientable 2-surfaces with constant Gaussian curvature, which include the static and spherically symmetric case. In this context, the GHP formalism proves to be a valuable tool to explore how terms in Eq. \eqref{ec:Smarr_Padmanabhan} are geometrically encoded on a trapping horizon, and particularly in the study of future outer marginal surfaces that foliate the trapping horizon.

Let us assume that the spacetime $(M,g)$ is defined as a $\mathcal{C}^k$-manifold and the metric tensor is of type $\mathcal{C}^{k-1}$. This assumption on differentiability allows the field equations of a extended theory of gravity of order $k-1$ to be well-defined everywhere on $M$. Through all the manuscript we use the signature (1,3) for the metric and the Penrose and Rindler's sign convention for the Riemann curvature tensor \cite{Penrose1984}.

We consider any modified theory of gravity whose associated field equations can be described as
\begin{equation}\label{ec:General_FieldEq}
-R_{\mu\nu} = \Delta t_{\mu\nu}
\end{equation}
where $\Delta t_{\mu\nu}$ are the components of some effective stress-energy tensor which does not include the Ricci tensor. Many general Lagrangians allow the field equations to take the form of Eq. \eqref{ec:General_FieldEq}. Some examples include GR, $f(R)$ gravity, scalar-tensor theories, etc. For the aforementioned examples we have: (i) within GR, the effective stress energy tensor is given by $\Delta t_{\mu\nu}^{(GR)} = T_{\mu\nu}^{(m)} - \frac{1}{2}T^{(m)}g_{\mu\nu} + \lambda g_{\mu\nu}$, where $T_{\mu\nu}^{(m)}$ is the stress-energy tensor associated with the material Lagrangian and $\lambda$ is the cosmological constant; (ii) the effective stress energy tensor in $f(R)$ gravity is given by
\begin{equation}\label{stress_energy}
     \Delta t_{\mu \nu} \equiv F(R)^{-1} \Big( -\frac{1}{2} f(R) g_{\mu \nu}+\big[\nabla_{\mu} \nabla_{\nu}-g_{\mu \nu} \,\Box] F(R) \Big),
\end{equation}
where $F(R) \coloneqq df(R)/dR$; and (iii) for scalar-tensor theories it holds that
\begin{equation}
    \Delta t_{\mu\nu}^{(\phi)} = \dfrac{1}{\phi} T_{\mu\nu}^{(m)} - \dfrac{1}{2}Rg_{\mu\nu} + \dfrac{1}{\phi}\left[\nabla_\mu\nabla_\nu-g_{\mu\nu}\square\right]\phi + \dfrac{\omega(\phi)}{\phi^2}\left(\partial_\mu\phi\partial_\nu\phi + \dfrac{1}{2}g_{\mu\nu}\partial_\sigma\phi\partial^\sigma\phi\right),
\end{equation}
where $\phi$ is an auxiliar (scalar) field which must be a solution of the corresponding scalar-field equation \cite{Quiros2019}.

To the best of our knowledge, there has been no comprehensive analysis investigating which classes of Lagrangian allow this specific formulation of the field equations. A systematic study in this direction might not only broaden our understanding of existing theoretical frameworks but also inspire the development of new theories of gravity.

In alignment with the spin coefficient and GHP formalisms, we construct a null tetrad consisting of two null vectors, $\mathbf{l}$ and $\mathbf{n}$. For the cases examined in this manuscript, these vectors can be directly aligned with the principal null directions \cite{NewmanPenrose}. The tetrad is completed by introducing a complex null vector $\mathbf{m}$ and its conjugate $\overline{\mathbf{m}}$, formed by combining two real, orthonormal spacelike vectors.

Within the spin coefficient formalism, and also in the GHP form, the so called Ricci scalars, $\Phi_{ij}$ with $i,j\in\{0,1,2\}$, are defined as the contractions of the Ricci tensor with the null tetrad. These are the components of the Ricci tensor in the new basis \cite{NewmanPenrose, GHPBook}. Thus, the corresponding field equations, Eq. \eqref{ec:General_FieldEq}, in this new basis can be written as $\Phi_{ij} = \Phi_{ij}^{ph}$ and $\Lambda = \Lambda^{ph}$, where $\Phi_{ij}^{ph}$ define the \textit{physical contractions} of the effective stress-energy tensor in an analogous way as the Ricci scalars and $\Lambda^{ph}=\Delta t/24$, where $\Delta t=\Delta t^{\mu}_{\mu}$ is the trace of the effective stress-energy tensor.

At this point, the field equations in the new basis can be rewritten in terms of geometrical quantities by using the GHP equations. Here, $\thorn$ (thorn), $\eth$ (edth) and their primed form denote the spin weighted derivatives, related with the directional derivatives along the null tetrad vectors, respectively, where primes ($'$) indicate transformations under the exchange $\mathbf{l} \leftrightarrow \mathbf{n}$ and $\mathbf{m} \leftrightarrow \overline{\mathbf{m}}$.

and the definition of the Penrose-Rindler $K$-curvature (see Eq.~(4.14.20) in \cite{Penrose1984}). That is, the field equations are 
\begin{eqnarray}
\Phi_{00}^{ph} & = & \thorn\rho - \eth'\kappa - \rho^2 - \sigma \overline{\sigma} + \overline{\kappa}\tau + \tau'\kappa,\\
2\Phi_{01}^{ph} & = & \thorn\tau - \thorn'\kappa + \eth\rho - \eth'\sigma - (\tau-\overline{\tau}')\rho - (\overline{\tau}-\tau')\sigma - (\rho-\overline{\rho})\tau - (\overline{\rho}'-\rho')\kappa,\hspace{0.5cm}\phantom{o}\\
\Phi_{11}^{ph}+3\Lambda^{ph} & = & K - \sigma\sigma' + \rho\rho' + (\Psi_2 + 2\Lambda),\label{ec:Phi11+3Lambda}
\end{eqnarray}
where overline denote complex conjugation and $\Psi_2$ is the Weyl curvature scalar
\begin{equation}
    \Psi_2 = C_{\alpha\beta\gamma\delta} l^{\alpha} m^{\beta} \overline{m}^{\gamma} n^{\delta}.
\end{equation}

Additional field equations are obtained by systematically applying prime ($'$), complex conjugate (c.c.) or Sachs ($\ast$) operators recursively as the diagrams in Fig. \ref{fig:diagrams} shows,
\begin{figure}[ht]
\begin{center}
\begin{tikzpicture}[every node/.style={minimum size=20pt, inner sep=2pt}]
    \node (A) at (0, 0) {$\mathbf{(0,0)}$};
    \node (B) at (2, 0) {$(0,2)$};
    \node (C) at (0, 2) {$(2,2)$};
    \node (D) at (2, 2) {$(2,0)$};

    \draw[<->] (A) -- (B);
    \draw[<->] (B) -- (D);
    \draw[<->] (D) -- (C);
    \draw[<->] (C) -- (A);

    \draw[->] (A) edge[loop left] node {c.c.} (A);
    \draw[->] (C) edge[loop left] node {c.c.} (C);

    \node at (1, -0.3) {${}^\ast$};
    \node at (-0.2, 1) {$'$};
    \node at (1, 2.2) {${}^\ast$};
    \node at (2.75, 0.95) {$'\text{ / c.c.}$};
\end{tikzpicture}
\end{center}
\vspace{0.5cm}

\begin{center}
\begin{tikzpicture}[every node/.style={minimum size=20pt, inner sep=2pt}]
    \node (A) at (0, 0) {$\mathbf{(0,1)}$};
    \node (B) at (2, 0) {$(2,1)$};
    \node (C) at (0, 2) {$(1,0)$};
    \node (D) at (2, 2) {$(1,2)$};

    \draw[<->] (A) -- (B);
    \draw[<->] (B) -- (D);
    \draw[<->] (D) -- (C);
    \draw[<->] (C) -- (A);

    \draw[->] (A) edge[loop left] node {$\ast$} (A);
    \draw[->] (B) edge[loop right] node {$\ast$} (B);

    \node at (1, -0.35) {$'$};
    \node at (-0.5, 1) {$\text{c.c.}$};
    \node at (1, 2.4) {$'\text{ / }\ast$};
    \node at (2.5, 0.95) {$\text{c.c.}$};
\end{tikzpicture}
\end{center}
\vspace{0.5cm}

\begin{center}
\begin{tikzpicture}[every node/.style={minimum size=20pt, inner sep=2pt}]
    \node (A) at (0, 0) {$\mathbf{(1,1)_{+3\Lambda}}$};
    \node (B) at (3, 0) {$(1,1)_{-3\Lambda}$};
    
    \draw[<->] (A) -- (B);

    \draw[->] (A) edge[loop below] node {$'\text{ / c.c.}$} (A);
    \draw[->] (B) edge[loop below] node {$'\text{ / c.c.}$} (B);

    \node at (1.5, -0.3) {${}^\ast$};
\end{tikzpicture}
\end{center}

\caption{Schematic diagram showing the relation between the field equations in terms of the prime, Sachs and complex conjugate operators.}
\label{fig:diagrams}
\end{figure}
\noindent where $(i,j)$ denotes $\Phi_{ij}$. For instance, $(1,1)_{\pm 3\Lambda}$ means $\Phi_{11}\pm 3\Lambda$. Here, the Sachs operator ($\ast$) is defined by the following transformations:
\begin{equation}
    \mathbf{l}^{\ast} = \mathbf{m},\quad \mathbf{n}^{\ast} = -\overline{\mathbf{m}}, \quad \mathbf{m}^{\ast} = -\mathbf{l},\quad \overline{\mathbf{m}}^{\ast} = \mathbf{n}
\end{equation}
For further details check references \cite{Penrose1984, Geroch1973, GHPBook}.

This form of the field equations is valid for any Petrov type. Moreover, in the particular case that all the Ricci scalars are real, it follows from the previous diagrams that there are at most five independent field equations
as $\Phi_{20}=\Phi_{02}$, $\Phi_{00}=\Phi_{22}$ \footnote{$\Phi_{00} = \Phi_{02}^{\ast} = \overline{\Phi_{20}}^{\ast} = \Phi_{20}^{\ast} = \Phi_{22}$.} and $\Phi_{01}=\Phi_{10}=\Phi_{12}=\Phi_{21}$ \footnote{$\Phi_{01}= \Phi_{01}^{\ast} = \overline{\Phi_{01}}^{\ast} = \Phi_{10}^{\ast} = \Phi_{12} =  \overline{\Phi_{21}} = \Phi_{21}$.}. Note that Bianchi identities in GHP formalism do not reduce, in general, any degree of freedom of the field equations, as they involve derivatives of the Ricci and Weyl scalars.

In the particular case of spacetimes that can be foliated by spatial 2-surfaces with constant Gaussian curvature (this does not hold for leaves with non-constant curvature), all the Ricci scalars are real and $\Phi_{01} = \Phi_{02} = 0$, which reduces the number of nontrivial field equations to three. If, additionally, the spacetime is static and the metric tensor is fully characterized by a single function, it also holds\footnote{The proof is by direct computation taking the line element $ds^2 = f(r) dt^2 - g(r) dr^2 - r^2 d\Omega^2$, where $d\Omega^2 = a(x,y)dx^2 + b(x,y) dy^2$ and $g(r) = f(r)^{-1}$. This includes the case where the marginal trapped surfaces have constant Gaussian curvature case.} that $\Phi_{00} = 0$, which reduces the problem to two independent field equations.

This is quite relevant for some examples in static and spherically symmetric spacetimes, where the metric is given by Eq. \eqref{metric:static_spherical}, as the spacetime is foliated by 2-spheres, which are compact orientable 2-surfaces of constant Gaussian curvature. In such cases, we must focus on Eq. \eqref{ec:Phi11+3Lambda} and its corresponding Sachs-transformed equation as they are the only nontrivial field equations, which are represented in the lower diagram in Fig. \ref{fig:diagrams}.

The term $\Psi_2 + 2\Lambda$ can be removed in terms of the spin coefficients by using the GHP equations
\begin{eqnarray}
\thorn'\rho - \eth'\tau & = & \rho\overline{\rho}'+\sigma\sigma'-\tau\overline{\tau} - \kappa\kappa'-(\Psi_2+2\Lambda),\label{ec:RemovePsi2+2L}\\
\thorn\rho' - \eth\tau' & = & \rho\overline{\rho}'+\sigma\sigma'- \tau'\overline{\tau}'-\kappa\kappa'-(\Psi_2+2\Lambda).
\end{eqnarray}

Based on Hayward's definition of trapping gravity \cite{Hayward1994},
\begin{equation}\label{def:hayward_trapping_grav}
\kappa = \sqrt{\dfrac{\thorn'\rho}{\chi\overline{\chi}}},
\end{equation}
which is the generalization of the surface gravity\footnote{Here, $\chi\overline{\chi}$ is a normalization term. Notice that Hayward assumes a gauge for the GHP formalism adapted to the double-null foliations of the spacetime with may be incompatible with $\chi=1$. Through this manuscript we choose the gauge $\chi = 1$.}, we choose\footnote{In fact, any other choice will lead to the same result in all the cases we are studying.} to rewrite the field Eq. \eqref{ec:Phi11+3Lambda} and its corresponding Sachs-transformed equation by using only Eq. \eqref{ec:RemovePsi2+2L}. Thus,
\begin{eqnarray}
K + \rho\rho' + \rho\overline{\rho}'-\tau\overline{\tau} - \kappa\kappa'-\thorn'\rho +\eth'\tau & = & \Phi_{11}^{ph}+3\Lambda^{ph},\label{ec:pre_kg}\\
K^* +\sigma\sigma' + \rho\overline{\rho}'-\tau\overline{\tau}- \tau\tau'-\thorn'\rho + \eth'\tau & = & -(\Phi_{11}^{ph}-3\Lambda^{ph}).\label{ec:pre_kg_sachs}
\end{eqnarray}

Moreover, recall that
\begin{equation}\label{ec:GaussianCruvature}
k_g = K + \overline{K}
\end{equation}
is the Gaussian curvature of the considered spacelike 2-surfaces (see Proposition~4.14.21 in \cite{Penrose1984} and App. \ref{app:foliations}). That is, $k_g$ is the Gaussian curvature of the future outer marginal surface when evaluating Eq. \eqref{ec:pre_kg} at the horizon\footnote{From now on, by horizon we are referring to the future outer marginal surfaces which foliates the trapping horizon in the Hayward's definition of black hole \cite{Hayward1994}. Do not confuse with the global definition of event horizon. Essentially, this is a cross section of the trapping horizon. This terminology is widely used in the literature.}. Using this equation, Eqs. \eqref{ec:pre_kg} and \eqref{ec:pre_kg_sachs} can be rewritten as
\begin{eqnarray}
k_g + 2\,\mathrm{Re}(\rho\rho') + 2\,\mathrm{Re}(\rho\overline{\rho}') - 2\tau\overline{\tau} - 2\,\mathrm{Re}(\kappa\kappa') -2\,\mathrm{Re}(\thorn'\rho) + 2\,\mathrm{Re}(\eth'\tau) & = & 2(\Phi_{11}^{ph}+3\Lambda^{ph}),\label{ThermoFieldEq:k_g}\\
\Tilde{k}_g + 2\,\mathrm{Re}(\sigma\sigma') -2\tau\overline{\tau} - 2\,\mathrm{Re}(\rho\overline{\rho}') - 2\,\mathrm{Re}(\tau\tau')-2\,\mathrm{Re}(\thorn'\rho) + 2\,\mathrm{Re}(\eth'\tau) & = & -2(\Phi_{11}^{ph}-3\Lambda^{ph}).\label{ThermoFieldEq:k_g_tilde}
\end{eqnarray}
where we defined $\Tilde{k}_g = K^* + \overline{K}^*$. At this point, we do not ascribe any geometric meaning to $\Tilde{k}_g$. For example, when the principal null directions of a Petrov D spacetime define a Lorentz surface, $\Tilde{k}_g$ can be interpreted as its Gaussian curvature. That is our case of study, see App. \ref{app:foliations}.

These equations, Eqs. \eqref{ThermoFieldEq:k_g} and \eqref{ThermoFieldEq:k_g_tilde}, when evaluated at a horizon, encode all the thermodynamic information in a precise way that will be developed below. This fact is one of the key points of this manuscript.

\subsection*{Geometric interpretation of black hole horizon}

After considering Eqs. \eqref{ThermoFieldEq:k_g} and \eqref{ThermoFieldEq:k_g_tilde} evaluated at the horizon, i.e. $\rho=0$, we arrive to  
\begin{equation}\label{ec:horizon_balance}
k_g - \Tilde{k}_g = 4\Phi_{11}^{ph} - 2\,\mathrm{Re}(\tau\tau') + 2\,\mathrm{Re}(\sigma\sigma').
\end{equation}

This allows a geometric interpretation of the definition of black hole horizon, characterized by
the equilibrium between material, rotation and shear energy densities on the right hand side and a surface tension term on the left hand side. This relation suggest that $\Tilde{k}_g$ has an interpretation of energy density and $K-K^*$ has dynamical meaning, which was pointed out by Hayward \cite{Hayward1994}.

In particular, for the vacuum case in static spacetimes foliated by 2-surfaces of constant Gaussian curvature, the horizon can be simply characterized as a balance between the curvatures, i.e. $k_g = \Tilde{k}_g$.


\section{\label{sec:main_findings}Smarr relation in GHP formalism}

We will assume that $(M,g)$ is a static spacetime which can be foliated by orientable Riemannian 2-surfaces of constant Gaussian curvature. This allow us to write the line element as
\begin{equation}
ds^2 = f(r) dt^2 - g(r) dr^2 - q(r) d\Omega^2.
\end{equation}
as is shown in App. \ref{app:foliations}.

The choice $q(r) = r_0^2$ leads to the existence of degenerated trapping horizons which do not correspond with the presence of a black hole, e.g. the Nariai spacetime. Thus, we restrict ourselves to the case where $q(r)$ has non-zero gradient. Then, there is always a suitable coordinate changes so $q(r)=r^2$ can be taken as the most general case. In particular, we are interested in cases where the Riemannian 2-surfaces are compact. This includes the spherically symmetric case or more exotic ones as the flat torus $T^2 = \mathbb{R}^2/\mathbb{Z}^2$ or Fuchsian surfaces $\mathbb{H}/\Gamma$ with genus $g\geq 2$, where $\mathbb{H}$ is the hyperbolic plane and $\Gamma$ is a Fuchsian group.

Following Padmanabhan's approach, we first study a simplified case taking $g(r) = f(r)^{-1}$. This choice of the metric function implies that $G^t_t = G^r_r$ (in fact, this is true for any static and spherically symmetric black hole horizon even if the metric is not determined by a single function \cite{Medved2004})
and then the only two independent field equations are $G^r_r = T^r_r$ and $G^{\theta}_{\theta} = T^{\theta}_{\theta}$.

In terms of the Ricci coefficients, this choice of the metric implies that the only non-vanishing components are $\Phi_{11}$ and $\Lambda$. Thus, these coefficients must be related with the different components of the Einstein tensor. As we showed in Sec. \ref{sec:general_case}, the only nontrivial field equations are the ones given in the latter diagram in Fig. \ref{fig:diagrams}. Interestingly,
\begin{eqnarray}
    \Phi_{11}+3\Lambda & = & -\frac{1}{2}G^r_r,\\
    \Phi_{11}-3\Lambda & = & \frac{1}{2}G^{\theta}_{\theta},
\end{eqnarray}
which allow reading these GHP Eqs. \eqref{ThermoFieldEq:k_g} and \eqref{ThermoFieldEq:k_g_tilde} as a thermodynamic relation.

In this particular case, the spacetime is of type D and the congruences associated to the principal null directions are geodesic, shear-free and non-rotating. Then $\kappa = \sigma = 0$ and $\omega = \mathrm{Im}(\rho) = 0$, as is shown in App. \ref{app:foliations}. This implies that $\rho$ is a real quantity that determines the congruence expansion of the principal null directions normal to the marginal surface \cite{GHPBook}. At this point, Eq. \eqref{ThermoFieldEq:k_g} reads
\begin{eqnarray}\label{ThermoEq:SphericalSym}
k_g - 2 \thorn'\rho = 2(\Phi_{11}^{ph}+3\Lambda^{ph}).
\end{eqnarray}

The first term in Eq. \eqref{ThermoEq:SphericalSym} is strongly related with the Hayward definition of quasi-local energy \cite{Hayward1994}
\begin{equation}
U = \dfrac{1}{4\pi}\sqrt{\dfrac{A}{16\pi}}\int_H \star k_g\, ,
\end{equation}
where $A$ denotes the area of the black hole horizon and $\star$ denotes the Hodge operator defined on the horizon $H$. This coincides with the energy term obtained by Padmanabhan, $E=a/2$, for the static and spherically symmetric case \cite{Padmanabhan2010}. 

Moreover, we give a generalized definition for the internal energy of a black hole following the Hayward's procedure \cite{Hayward1994b}. Over a compact 2-surface $S$, Hayward introduces the Hamiltonian 2-form
\begin{equation}
    \mathcal{H} = \star\dfrac{1}{8\pi}\left(K + \rho\rho'-\sigma\sigma'+\tau\tau'\right),
\end{equation}
which is the corresponding dynamical object for the Einstein field. The quantity $\int_H \mathcal{H}$ must be multiplied by some length in order to have energy units. This motivates a generalized definition of black hole internal energy valid for exotic black hole topologies given by\footnote{Note that the term $\rho\rho'$ is zero over the horizon and $\sigma\sigma' = \tau\tau' = 0$ for this metric ansatz.}
\begin{equation}\label{def:generalized_energy}
U = \dfrac{1}{8\pi}\sqrt{\dfrac{A}{A_0}}\int_H\,\star k_g,
\end{equation}
where $A_0$ is a dimensionless constant representing the area of a \textit{unit horizon}\footnote{We use $\sqrt{A/A_0}$ as it is the most natural unit of length.}, being $A_0 = 4\pi \vert g-1\vert$ in the non-flat case (where $g$ is the genus of the horizon) and $A_0 = L_x L_y$ in the flat torus case\footnote{This is the unique compact surface of constant Gaussian curvature equal to zero.}, where $L_x$ and $L_y$ are the length of the coordinate ranges after the identifications $x \sim x + L_x$ and $y \sim y+ L_y$. In particular, introducing the Euler characteristic of the horizon, denoted by $\chi(H)$, Eq. \eqref{def:generalized_energy} can be written as $U = \frac{\chi(H)}{2}\left(\frac{a}{2}\right)$ which coincides with the Padmanabhan's definition of internal energy \cite{Padmanabhan2010}.

The quantity $\thorn'\rho$ is constant over a stationary horizon \cite{Hayward1994d}. Moreover, by direct computation $\Phi_{11}^{ph}+3\Lambda^{ph}$ depends only on the $r$ coordinate and the Gaussian curvature of the marginal surface, which we assumed to be constant. Thus, after using Eq. \eqref{def:generalized_energy} in Eq. \eqref{ThermoEq:SphericalSym}, this can be rewritten as
\begin{equation}\label{ec:hodge_integrated}
\frac{\chi(H)}{4} \sqrt{\dfrac{A}{A_0}} - \dfrac{1}{4\pi}\sqrt{\dfrac{A}{A_0}} \thorn'\rho\, A = \dfrac{1}{4\pi}\sqrt{\dfrac{A}{A_0}}(\Phi_{11}^{ph}+3\Lambda^{ph}) A,
\end{equation}
which will be referred as the Penrose-Rindler Smarr formula. Here, we can identify the first term on the previous relation as the total internal energy of the black hole
\begin{equation}
U = \frac{\chi(H)}{4} \sqrt{\dfrac{A}{A_0}},
\end{equation}
which, again, coincides with Hayward and Padmanabhan's definitions in the spherically symmetric case.

Moreover, the key feature of this Penrose-Rindler Smarr formula is that it motivates a revised definition of black hole temperature given by
\begin{equation}\label{def:temperature}
T = \dfrac{\hbar}{2\pi} \sqrt{\frac{A}{A_0}}\thorn'\rho,
\end{equation}
after identifying the Bekenstein-Hawking entropy $S = A/4\hbar$ in the GR case. This definition is a scalar invariant of the foliation and lead to the following redefinition of trapping gravity
\begin{equation}\label{def:trapping_gravity}
\kappa = \sqrt{\dfrac{A}{A_0}} \thorn'\rho.
\end{equation}

In particular, for the spherically symmetric case, this definition gives $\kappa = \sqrt{\frac{A}{4\pi}}\thorn'\rho$, which coincides with the surface gravity.

As we commented before, Hayward's definition for trapping gravity \cite{Hayward1994}, shown in Eq. \eqref{def:hayward_trapping_grav} does not coincide with the surface gravity even in the context of static and spherically symmetric spacetimes , e.g. the Reissner-Nördstrom solution \cite{Fodor1996,Nielsen2008}. Our definition in Eq. \eqref{def:trapping_gravity} overcomes this problem and is valid for any static spacetime foliated by compact orientable 2-surfaces with constant Gaussian curvature.

Furthermore, there is an additional pressure term related to the matter content present in the spacetime as
\begin{equation}\label{def:pressure_term}
P = -\dfrac{1}{4\pi}(\Phi_{11}^{ph}+3\Lambda^{ph}),
\end{equation}
which coincides with the Padmanabhan's definition\footnote{Take into account the $8\pi$ factor due to the definition of the stress-energy tensor.} of pressure in GR. 

At this point, some comments about the thermodynamic volume of the black hole should be made. We first analyze the vacuum case in GR, where $\Phi_{11}^{ph} = 0$, which admits a cosmological constant. Such a case has been widely studied, identifying the presence of a cosmological constant with a $P\mathrm{d}V$ term in the first law of black hole thermodynamics \cite{Kubizk2017}. Indeed, this allows the definition of the thermodynamic volume of the black hole, which may not coincide with the volume of a 3-manifold defined as having the horizon as its boundary (this 3-manifold may even be not well defined, for example in some hyperbolic cases). Thus, the thermodynamic volume should be defined as the conjugate thermodynamic variable to the pressure
\begin{equation}
V = \left(\dfrac{\partial M}{\partial P}\right)_S.
\end{equation}

As commented in Sec. \ref{sec:rep_padma}, when a cosmological constant is considered, the mass of the black hole no longer coincides with its internal energy, but it is identified with the chemical enthalpy \cite{Kubizk2017}. This is,
\begin{equation}\label{def:chemical_enthalpy}
M = U + P_{\Lambda}V.
\end{equation}

This motivates the definition of the \textit{thermodynamic volume} of the black hole as
\begin{equation}
V = \sqrt{\dfrac{A}{A_0}}\left(\dfrac{A}{3}\right)
\end{equation}
which coincides with the volume of the 2-sphere in the spherically symmetric case,
\begin{eqnarray}
    V = \dfrac{4}{3}\pi a^3.
\end{eqnarray}

Considering the aforementioned volume, we are able to write the Penrose-Rindler Smarr formula in the same terms as Padmanabhan's result. That is,
\begin{equation}
U = 2TS - 3PV.
\end{equation}

Again, when the pressure term is only due to a negative cosmological constant (and $\Phi_{11}^{ph} = 0$) then, by using Eq. \eqref{def:chemical_enthalpy}, Eq. \eqref{ec:hodge_integrated} can be written as
\begin{equation}
M = 2TS - 2P_{\Lambda}V,
\end{equation}
which let us identify the Penrose-Rindler Smarr formula as the Smarr relation.

Moreover, in the $\Phi_{11}^{ph} = 0$ case, there is a one-to-one correspondence between geometric quantities and \textit{physical} quantities: relating the Gaussian curvature with the internal energy, the rate of change of the expansion along a principal null direction with the temperature, and the Ricci scalar with a cosmological pressure in the AdS case.

However, when returning to the general case with $\Phi_{11}\neq 0$, the resulting Penrose-Rindler Smarr formula is again the Smarr relation, but the Riemann decomposition is not the \textit{physical} decomposition as the mass of the black hole is not $M=U+PV$ when $P$ is defined as in Eq. \eqref{def:pressure_term}. An explicit example is shown below. Despite this, if the spacetime is stationary and asymptotically flat, the term $U+3PV$ coincides with the Komar energy of the spacetime \cite{Banerjee2010}.

Finally, as an aside note, see that Eq. \eqref{ThermoFieldEq:k_g_tilde} is equivalent to Eq. \eqref{ThermoEq:SphericalSym} as the black hole horizon is determined by $k_g - \Tilde{k}_g = 4\Phi_{11}^{ph}$ using Eq. \eqref{ec:horizon_balance}. That is, the equation $G^{\theta}_{\theta} = T^{\theta}_{\theta}$ can also be read as a thermodynamic relation. Additionally, this procedure can be generalized to the case $\Phi_{00}\neq 0$ as the thermodynamic information remains in Eqs. \eqref{ThermoFieldEq:k_g} and \eqref{ThermoFieldEq:k_g_tilde}.

To illustrate these results, we consider two specific examples: (i) within GR, we employ the GHP formalism to reobtain the Smarr formula for the Reissner-Nordström-AdS black hole and (ii) we adapt the Penrose-Rindler Smarr formula to $f(R)$ gravity.

\subsection*{Reissner-Nördstrom black hole in AdS background: a GHP derivation}

The Reissner–Nordström black hole is a static solution to Einstein's field equations, describing a non-rotating, spherically symmetric black hole that possesses both mass and electric charge. This type of black hole is characterized by two different horizons: an outer horizon similar to the Schwarzschild horizon, which plays the role of the event horizon, and an inner horizon related to its charge, which is a Cauchy horizon. Physically, the inner horizon marks the boundary of predictability. Beyond this horizon, the usual initial conditions required to predict the future evolution (based on Einstein’s equations) no longer hold, meaning the spacetime geometry may be fundamentally unpredictable. 

When the electric charge reaches a critical value relative to the mass, these horizons converge, creating an \textit{extremal} black hole with unique properties, such as zero Hawking temperature and heightened stability against certain perturbations. Interestingly, this feature is captured in the revised temperature definition given in this manuscript, as in this extremal case $\thorn'\rho = 0$ holds.

Now, we show that the Penrose-Rindler Smarr formula is exactly reduced to the Smarr formula for the non-extremal Reissner-Nördstrom black hole when rearranging terms. Recall that we were able to write Eq. \eqref{ec:hodge_integrated} as
\begin{equation}\label{ec:geom_smarr_form}
U = 2TS - 3PV,
\end{equation}
where $P = -\frac{1}{4\pi}(\Phi_{11}^{ph}+3\Lambda^{ph})$ is the pressure term and $V$ is the thermodynamic volume.

For the Reissner-Nördstrom black hole, the term $\Phi_{11}^{ph}$ is given by
\begin{eqnarray}
\Phi_{11}^{ph} = \dfrac{Q}{2r^4},
\end{eqnarray}
with $Q$ being the black hole charge. Moreover, considering a negative cosmological constant $\lambda$, then $\Lambda^{ph} = \lambda/6$. Thus, Eq. \eqref{ec:geom_smarr_form} takes the form
\begin{equation}
U = 2TS + \dfrac{1}{2}Q\Phi - 3 P_{\lambda} V,
\end{equation}
where $\Phi = Q/r$ is the electric potential and $P_{\lambda} = -\lambda/8\pi$ the cosmological pressure. This expression can be rearranged as
\begin{eqnarray}
U + \dfrac{1}{2}Q\Phi + P_{\lambda} V = 2TS + Q\Phi - 2 P_{\lambda} V,
\end{eqnarray}
and noting that the left hand side term is equal to the black hole mass \cite{Villalba2020} this leads to
\begin{eqnarray}
M = 2TS + Q\Phi - 2P_{\lambda} V,
\end{eqnarray}
which is the Smarr relation for the Reissner-Nördstrom black hole in an AdS background.

Consequently, despite the geometric decomposition is exactly the \textit{physical} decomposition for the non-charged case, the Penrose-Rindler Smarr formula is not the \textit{physical} decomposition present in the Smarr formula. Again, all the information present in the Smarr formula is contained in Eq. \eqref{ec:GaussianCruvature}, but the correspondence between the $K$-curvature and physical quantities may not be one-to-one.

\subsection*{Smarr relation in $f(R)$ gravity: a GHP derivation}
For the vacuum case, the general action in $f(R)$ formalism is \cite{CapozzielloBEG}
\begin{equation}
S[g_{\mu \nu}] = \int d^4x \sqrt{-g}f(R),
\end{equation}
where we have set $16 \pi G=1$, with $G$ denoting the gravitational constant.

Applying the variational principle, $\delta S=0$, the field equations for $f(R)$ gravity can be written as Eq. \eqref{ec:General_FieldEq}, where we have defined the effective stress-energy tensor as in Eq. \eqref{stress_energy}.

The Smarr relation can be extended to $f(R)$ gravity by multiplying the Penrose-Rindler Smarr formula by the adimensional quantity $F(R)$, obtaining
\begin{equation}
\dfrac{\chi(H)}{4}\sqrt{\dfrac{A}{A_0}}F(R) - \dfrac{1}{4\pi}\sqrt{\dfrac{A}{A_0}}\thorn'\rho A F(R) = \dfrac{3}{4\pi}\left(\Phi_{11}+3\Lambda\right)^{ph} F(R) V.
\end{equation}

The first term coincides with the internal energy of the black hole in $f(R)$ gravity for the spherically symmetric case \cite{Sharif2012}
\begin{equation}
U = \dfrac{\chi(H)}{4}\sqrt{\dfrac{A}{A_0}}F(R).
\end{equation}

Additionally, the temperature remains the same independently of the theory (this is a expected result as the temperature is entirely determined by the geometry of the horizon) after identifying the corresponding entropy \cite{Faraoni2010}
\begin{equation}
S = \dfrac{A F(R)}{4\hbar}.
\end{equation}

Finally, note that there is an addition $F(R)$ term in the part that takes the role of a pressure that must be identified in each particular case.

\section{Towards the thermodynamics of more general configurations}

Despite we introduced the Smarr formula within the GHP formalism for static spacetimes foliated by compact spacelike 2-surfaces of constant Gaussian curvature, the linking between Penrose and Rindler $K$-curvature and the Smarr formula suggests that the relation introduced in Eq. \eqref{ThermoFieldEq:k_g} is still related with the Smarr relation when integrated over a marginal surface for more general configurations. In this section, we briefly analyze\footnote{We thank the anonymous referee for pointing this out.} the encoding of the Smarr formula in the Penrose and Rindler $K$-curvature for some examples that are not contemplated in the initial spacetime geometric assumptions.

In particular, we study the charged Taub–NUT–AdS spacetime. The uncharged case is characterized by an extra parameter, $n$, which plays a role analogous to a \textit{gravomagnetic} charge. In the metric there is a crossed term combining the time coordinate together with an angular one, reflecting a frame-dragging effect of the spacetime and the presence of so-called Misner strings singularities extending along the polar axes. For a detailed introduction see \cite{HawkingEllis}.

The charged Taub–NUT–AdS spacetime is a solution of Einstein–Maxwell gravity that generalizes the Taub–NUT–AdS geometry by incorporating electric and magnetic charges \cite{Taub1951,Newman1963}. In such a case, the line element is given by
\begin{equation}
    ds^2 = f(r)\left(dt+2n\cos\theta\, d\phi\right)^2 -f(r)^{-1}dr^2 - (r^2+n^2)\left(d\theta^2 +\sin^2\theta\, d\phi^2\right),
\end{equation}
with
\begin{equation}
    f(r) = \dfrac{r^2-2m r -n^2 +4n^2g^2+e^2}{r^2+n^2} -\dfrac{3n^4 - 6n^2r^2-r^4}{l^2(r^2+n^2)},
\end{equation}
where $e$ is the elerctric charge, $g$ is the magnetic charge and $l$ is the AdS radius.

In contrast with the ansatz introduced in Sec. \ref{sec:main_findings}, the principal null directions are rotating but again geodesic and shear-free, which implies $\omega\neq 0$ and $\kappa = \sigma = 0$. Recalling that Eq. \eqref{ThermoFieldEq:k_g} is a general relation valid for any spacetime, which is simplified to
\begin{equation}
    k_g - 2\tau\overline{\tau} - 2\,\mathrm{Re}(\kappa\kappa') -2\,\mathrm{Re}(\thorn'\rho) + 2\,\mathrm{Re}(\eth'\tau) = 2(\Phi_{11}^{ph}+3\Lambda^{ph}),
\end{equation}
when evaluated at the horizon, thus it leads to the expression
\begin{equation}
    k_g -2\,\mathrm{Re}(\thorn'\rho) + 2\,\mathrm{Re}(\eth'\tau) = 2(\Phi_{11}^{ph}+3\Lambda^{ph}),
\end{equation}
as $\tau =0$ for charged Taub-NUT AdS spacetimes. Hence, when integrating over the marginal surface we obtain
\begin{equation}\label{eq:general_TaubNUT_rel}
    2\pi \chi(H) - 2\,\mathrm{Re}(\thorn'\rho) A = 2(\Phi_{11}^{ph}+3\Lambda^{ph}) A,
\end{equation}
as $\thorn'\rho, \Phi_{11}^{ph}$ and $\Lambda^{ph}$ are constant over the marginal surfaces and $\eth'\tau$ is a gradient which has zero integral \cite{Hawking1972}. 

Following a similar discussion that the one introduced in Sec. \ref{sec:main_findings}, Eq. \eqref{eq:general_TaubNUT_rel} cannot be identified with the Smarr formula as it does not have energy units. Therefore, it is necessary to multiply the previous relation by a length dimension. 

In contrast with the previous examples, the finding problem in this case is that Eq. \eqref{def:temperature} does not correctly defines the temperature of a charged Taub-NUT AdS black hole. This result suggests that the factor $\sqrt{A/A_0}$ is not valid for such a general case. Despite this, the Smarr formula can be recovered by multiplying Eq. \eqref{eq:general_TaubNUT_rel} by $\frac{1}{8\pi}\frac{r_+^2+n^2}{r_+}$ (that is, we are replacing $\sqrt{A/A_0}\to\frac{r_+^2+n^2}{r_+}$), leading to
\begin{equation}\label{multiplied_relation}
    \frac{\chi(H)}{4}\frac{r_+^2+n^2}{r_+} - \frac{1}{4\pi}\frac{r_+^2+n^2}{r_+}\mathrm{Re} (\thorn'\rho)\,A = \frac{1}{4\pi}\frac{r_+^2+n^2}{r_+}(\Phi_{11}^{ph}+3\Lambda^{ph})A.
\end{equation}

The geometric motivation and interpretation of the factor $\frac{r_+^2+n^2}{r_+}$ is beyond the scope of this manuscript and it motivates an interesting path for linking Eq. \eqref{ThermoFieldEq:k_g} with the Smarr formula in even more generic configurations.

Note that $\chi(H)=2$ as the cross sections of the horizon are topological 2-spheres. Thus, the first term in Eq. \eqref{multiplied_relation} is given by $\frac{r_+^2+n^2}{2r_+}$, which coincides with the black hole mass for the case where no cosmological constant is considered \cite{Liu2024}. In such a case, as previously introduced in the manuscript, the mass is identified with the internal energy leading to
\begin{equation}
    U = \dfrac{r_+^2+n^2}{2r_+}.
\end{equation}

Moreover, the black hole temperature can be identified as
\begin{equation}
    T = \dfrac{\hbar}{2\pi}\dfrac{r_+^2+n^2}{r_+}\mathrm{Re}(\thorn'\rho),
\end{equation}
after identifying the Bekenstein-Hawking entropy term $S = \frac{A}{4\hbar}$. In particular, a direct computation reveals that
\begin{equation}\label{decomposed_id}
    U = 2TS + \dfrac{1}{4\pi}(\Phi_{11}^{ph}+3\Lambda^{ph})A,
\end{equation}
which implies that Eq. \eqref{multiplied_relation} is identically to
\begin{equation}
    M=2(TS+\psi N - PV)+\phi_eQ_e + \phi_mQ_m,
\end{equation}
where all the 
well-known thermodynamic variables are described in \cite{Bordo2020,Hennigar2019}.

That is, we just showed that Eq. \eqref{ThermoFieldEq:k_g} is again related with the Smarr formula even in more general ansatz, although finding the necessary physical decomposition can be quite tedious for each particular case.


\section{\label{sec:final_remarks}Final remarks and comments}

In this manuscript we have studied how black hole thermodynamics is encoded in the field equations in terms of geometrical quantities as a result of the decomposition of the Riemann tensor using the GHP formalism. By giving a revised definition for black hole trapping gravity, which address limitations of Hayward's initial definition for static and spherically symmetric cases, we have found that the Penrose and Rindler $K$-curvature is not just a geometric relation, but it coincides with Smarr relation when integrated over a horizon. Specifically, for General Relativity we have obtained the following correspondence,
\begin{equation}\label{equivalence}
    k_g = K+\overline{K}\Leftrightarrow M = 2TS + \Phi Q- 2PV.
\end{equation}

Note that it is possible to write a similar equivalence for extended theories of gravity once the corresponding field equations can be written $-R_{\mu\nu} = \Delta t_{\mu\nu}$, where $\Delta t_{\mu\nu}$ are the components of some effective stress-energy tensor which does not include the Ricci tensor. For example, we have checked the aforementioned equivalence for black holes in $f(R)$ gravity.

Even more, this framework allow us to study exotic topologies, obtaining some constraints on black hole existence in particular scenarios. For instance, the Penrose-Rindler Smarr formula implies that only spherical black holes with non-zero temperature can exist in General Relativity under the dominant energy condition. This recovers the well-known topology law for black holes \cite{Hayward1994b}. Furthermore, the generalized definition of internal energy reveals that flat toroidal black holes have zero internal energy, while some hyperbolic black holes exhibit negative internal energy, which has been pointed out in previous works \cite{Villalba2020}.

The GHP formalism was initially developed for four-dimensional spacetimes, but it has been recently generalized to higher dimensions \cite{Durkee2010}. A key distinction in higher-dimensional analyses is the absence of a Gauss-Bonnet theorem, which plays a crucial role in relating black hole geometry to the Smarr law. Nevertheless, this extension suggests the possibility of analyzing thermodynamic formulations in higher-dimensional settings.

It is important to remark that the temperature we have proposed is not valid for the stationary axisymmetric case. In fact, despite the surface gravity over the Kerr black hole horizon must be constant due to the horizon topology, the quantity $\thorn'\rho$ is not constant over the horizon (as it depends on angular coordinates), which suggests a revised generalization of trapping gravity for stationary spacetimes. A similar issue arises in the outlined charged Taub-NUT AdS case, where the presence of NUT charge further complicates the decomposition of relevant horizon quantities into the physical quantities. The study of this procedure applied to a more general ansatz is deferred to future work, including the connection between the Penrose and Rindler formula and the Smarr formula in Kerr-Newmann black holes or even in more generic configurations.


\section*{Acknowledgments}
We acknowledge financial support from the Generalitat Valenciana through PROMETEO PROJECT CIPROM/2022/13. A. G. acknowledges Fundación Humanismo y Ciencia for financial support. A.G. acknowledges Martuan for continuous support. We acknowledge the anonymous referees for their valuable comments and suggestions, which helped to improve this work.

\newpage

\appendix

\section{\label{app:foliations}Static spacetimes foliated by 2-surfaces of constant Gaussian curvature}

Suppose that there exist a foliation of $M$, denoted as $\mathcal{S}$, by Riemannian 2-surfaces. Let $\{\mathbf{e}_2,\mathbf{e}_3\}$ be an orthogonal basis of $T_p S$ for $S\in\mathcal{S}$.

Let $\mathbf{u}\in T_p M$ be a timelike unit vector normal to each leaf $S\in\mathcal{S}$, this is $\mathbf{u}\cdot d i_p (\mathbf{w}) = 0$ for all $\mathbf{w}\in T_p S \subset T_p M$, where $i : S \xhookrightarrow{} M$ denotes the inclusion map. The vector subspace $D_p = \text{span}(\mathbf{u})\oplus T_p S \subset T_p M$ is involutive by definition, as the Lie brackets of the subspace basis remain in $D_p$.

In that case, the Frobenius theorem assures that the distribution $D = \cup_{p\in M} D_p$ defines a foliation $\mathcal{F}$ of $M$, where the leaves of the foliation, $F\in\mathcal{F}$, are the hypersurfaces defined as the integral manifolds generated by $D_p$. Moreover, each leaf is timelike and $T_p F = D_p$ by construction. Thus, $\mathbf{u}$ defines an observer field on $M$.

At this point, it is possible to define a spacelike unit vector $\mathbf{v}\in D_p^{\perp}$ by using the Gram-Schmidt decomposition algorithm. Then, the vector subspace $E_p = \text{span}\{\mathbf{v}\}\oplus T_p S \subset T_p M$ is again involutive and the distribution $E = \cup_{p\in M} E_p$ defines a foliation $\mathcal{G}$ of $M$ by spacelike hypersurfaces, $G\in\mathcal{G}$, satisfying $T_p G = E_p$. Therefore, the set $\{\mathbf{u},\mathbf{v},\mathbf{e}_2,\mathbf{e}_3\}$ determine a orthogonal basis of $T_p M$.

Any foliation of $M$ can be written (locally) as a union set of level sets of a $\mathcal{C}^3$ function $\Phi : U\subset M \to \mathbb{R}^{d}$, with $d$ being the codimension of the leaves. Moreover, the vector $\Phi^{,\alpha}\mathbf{e_{\alpha}}$ is by definition normal to each leaf, where $\mathbf{e}_0 = \mathbf{u}$ and $\mathbf{e}_1 = \mathbf{v}$.

In this case we denote $t$ and $r$ the scalar functions which determine the foliations with $t^{,\alpha} \sim u^{\alpha}$ (timelike, normal to spacelike leaves $F\in\mathcal{F}$) and $r^{,\alpha} \sim u^{\alpha}$ (spacelike, normal to timelike leaves $G \in \mathcal{G}$). For this choice of coordinates, the line element induced by the metric tensor can be written as
\begin{equation}\label{def:initial_line_element}
ds^2 = g_{tt} dt^2 - g_{rr}dr^2 - ds_{S}^2.
\end{equation}

In our particular case, we impose that the surfaces $S\in\mathcal{S}$ are orientable compact surfaces of constant Gaussian curvature. Then, the surfaces $S\in\mathcal{S}$ are locally conformal to: (i) a sphere, (ii) a flat surface or (iii) a hyperbolic surface. Despite this is not explicitly commented along the manuscript, we consider that for constant coordinates of the parametrized 2-surfaces $S\in\mathcal{S}$, the resulting Lorentz surface at each point, with induced metric tensor $ds^2 = g_{tt}dt^2 - g_{rr}dr^2$, is invariant under infinitesimal variations along the directions $\mathbf{e}_2$ and $\mathbf{e}_3$. Consequently, assuming that the spacetime is static, the line element given in Eq. \eqref{def:initial_line_element} can be written as
\begin{equation}\label{def:line_element_constantKg}
ds^2 = f(r) dt^2 - g(r)dr^2 - q(r) d\Omega^2
\end{equation}
where $d\Omega^2$ denotes the line element of a spherical, planar or hyperbolic Riemannian surface, depending on whether the Gaussian curvature is taken positive, null or zero.

At this point, we define the null vectors
\begin{eqnarray}\label{nullbasis}
\mathbf{l} & = & \dfrac{1}{\sqrt{2}}(\mathbf{u}+\mathbf{v})\nonumber\\
\mathbf{n} & = & \dfrac{1}{\sqrt{2}}(\mathbf{u}-\mathbf{v})\\
\mathbf{m} & = & \dfrac{1}{\sqrt{2}}(\mathbf{e}_2+i\mathbf{e}_3)\nonumber
\end{eqnarray}
which, together with $\overline{\mathbf{m}}$, form a null tetrad and a basis of the complexified tangent space $T_p M^{\mathbb{C}}$. Moreover, the vectors $\mathbf{m}$ and $\overline{\mathbf{m}}$ define a basis of $T_p S^{\mathbb{C}}$.

It can be shown that, in this case, the vectors $\mathbf{l}$ and $\mathbf{n}$ define a dual-null foliation by taking $\tilde{D}_p = \text{span}(\mathbf{l})\oplus T_p S^{\mathbb{C}}$ and $\tilde{E}_p = \text{span}(\mathbf{n})\oplus T_p S^{\mathbb{C}}$ and proving that they are involutive. Moreover, it is shown by direct computation that $\mathbf{l}$ and $\mathbf{n}$ are the null principal directions of the Weyl tensor (up to a scaling factor), the spacetime is of type D and the marginal surfaces that foliate the trapping horizon (if it exists) are conformal to any $S\in\mathcal{S}$. Indeed, by using Eq. \eqref{nullbasis}, it can be shown that $\mathbf{l}$ and $\mathbf{n}$ define a Lorentz surface by constructing an involutive distribution generated by $\text{span}(\mathbf{l})\oplus\text{span}(\mathbf{n})$.

Finally, we should point out that the Gaussian curvature $k_g$, related to the Pernose-Rindler $K$-curvature by Eq. \eqref{ec:GaussianCruvature}, describes the curvature of the 2-surface orthogonal to $\mathbf{l}$ and $\mathbf{n}$. This is, $k_g$ is the Gaussian curvature of $S\in\mathcal{S}$.

\bibliographystyle{main}
\bibliography{main}

\end{document}